# SNC: A Cloud Service Platform for Symbolic-Numeric Computation using Just-In-Time Compilation

Peng Zhang[1], *Member, IEEE*, Yueming Liu[1], and Meikang Qiu, *Senior Member, IEEE*

*Abstract*— **Cloud services have been widely employed in IT industry and scientific research. By using Cloud services users can move computing tasks and data away from local computers to remote datacenters. By accessing Internet-based services over lightweight and mobile devices, users deploy diversified Cloud applications on powerful machines. The key drivers towards this paradigm for the scientific computing field include the substantial computing capacity, on-demand provisioning and cross-platform interoperability. To fully harness the Cloud services for scientific computing, however, we need to design an application-specific platform to help the users efficiently migrate their applications. In this, we propose a Cloud service platform for symbolic-numeric computation– SNC. SNC allows the Cloud users to describe tasks as symbolic expressions through C/C++, Python, Java APIs and SNC script. Just-In-Time (JIT) compilation through using LLVM/JVM is used to compile the user code to the machine code.**

**We implemented the SNC design and tested a wide range of symbolic-numeric computation applications (including nonlinear minimization, Monte Carlo integration, finite element assembly and multibody dynamics) on several popular cloud platforms (including the Google Compute Engine, Amazon EC2, Microsoft Azure, Rackspace, HP Helion and VMWare vCloud). These results demonstrate that our approach can work across multiple cloud platforms, support different languages and significantly improve the performance of symbolic-numeric computation using cloud platforms. This offered a way to stimulate the need for using the cloud computing for the symbolic-numeric computation in the field of scientific research.**

## I. INTRODUCTION

Cloud computing is becoming a prevailing provision of computing infrastructures for the enterprise and academic institutes towards enjoying a multitude of benefits: on-demand high performance computing capacity, location independent data storage, and high quality services access [1-3]. As usual, Cloud users can access to cloud services through Internet-based interfaces and Clouds offer the source provision "as a service". For examples, Cloud providers offer Infrastructure as a Service (IaaS) that provides virtualized computing resources that can host the users' applications and handle the associated tasks. Examples of IaaS providers include the Amazon Web Services (AWS), Microsoft's Windows Azure and Google Computer Engine. In addition, Platform as a Service (PaaS) offers the platform that allows the cloud users to develop and manage the underlying software. Google App Engine is one example of PaaS that is geared with a variety of tools such as Python, Java and SQL. Other types of Cloud services include: Software as a Service (SaaS) in which the software is licensed and hosted in Clouds; and Database as a Service (DBaaS) in which managed database services are hosted in Clouds. While enjoying these great services, we must face the challenges raised up by these unexploited opportunities within a Cloud environment.

Complex scientific computing applications are being widely studied within the emerging Cloud-based services environment [4-12]. Traditionally, the focus is given to the parallel scientific HPC (high performance computing) applications [6, 12, 13], where substantial effort has been given to the integration of numerical models with the computing facilities provisioned by a Cloud platform. Of using Cloud computing, the benefits include the dynamic provision: computing resources can be dynamically released as long as they are no more needed [7]; and the virtualization: the users are allowed to dynamically build their specific-purpose virtual clusters [4] and virtual workspace [5]. Thus, resource sharing and virtualization of the Cloud promise to efficiently offer on-demand computing services for demanding scientific computing [9-11]. In these cases, the Cloud platform demonstrated an alternative choice for the scientific community of classical scientific computing workloads. However, the obstacle to attracting scientific applications is a creation of a user-friendly environment where complex numeric algorithms could be efficiently programmed and deployed on the heterogeneous Cloud platforms. For the scientific field, many easy-to-use programming environments including Matlab, Maple, and Mathematica are expediting our research work. For example, they are offering the powerful symbolic tools for a wide range of scientific computing cases: numeric integration and differentiation of multivariable functions, solving the algebraic equations, optimizing nonlinear systems using symbolic derivatives, and computing integral transforms. These capabilities and functionalities are competitive and essential for completion of sophisticated scientific problems [14-16]. Basically, the users only need to describe sophisticated symbolic expressions in the high level environment, and leave the rest of massive calculations to the powerful toolboxes. As such, the toolboxes help scientific researchers rapidly program their ideas. Such programmability is fundamental to the progress and realization of the Cloud for scientific community. For example, several newly developed numerical methods for solving the inverse problems of partial differential equations [17-19] involve significant amount of symbolic manipulations of the mathematical expressions and numerical computations against the resulting expressions. The symbolic-numeric solution of modern scientific problems is becoming more and more popular. The 2015 J.H. Wilkinson Prize for numerical software was awarded to the authors of dolfin-adjoint [20], a package which automatically derives and solves adjoint and tangent linear equations from high-level mathematical specifications of the finite element discretization of PDE. In this, the core underlying library FEniCS utilized the way of source code generation for numerical computation. In this work, dynamic translation from the symbolic expressions to the machine codes is favored [21, 22]. In this regard, the just-in-time (JIT) compilation is a preferred approach than the source code generation and ahead-of-time (AOT) compilation.

---

[1] The authors contributed equally to this work.

To fully harness the Cloud resources for scientific computing services, it is by no means a trivial task. The challenges are at least three-fold: cross-language programming, cross-platform interoperability and on-demand provisioning. C/C++, Python and Java are now the dominating programming languages and are widely used by researchers. To lower the learning curves, a new platform must favor these popular languages. Just-in-time (JIT) compilation offers an opportunity of efficient interactive programming designs; however, it may also cause the difficulty of hosting dynamic programs within a Cloud environment. The heterogeneity of the hardware and software stacks may further escalate this challenge. Lastly, a scheduler is needed to handle the Internet-based user requests, produce the tasks, and schedule these tasks in Cloud. In this work, the SNC platform addressed these challenges and it offered the Cloud service for symbolic-numeric computation.

The main contributions include:

• SNC is the first platform to integrate symbolic-numeric computation with JIT compiler and offers cloud services. It features extremely fast numerical evaluation of symbolic expressions, zero administration and it is easy to scale.

• Proof-of-concept software is developed to demonstrate the applicability of design and it shows the efficacy and efficiency of SNC.

• The multi-language support and cross-platform capability are demonstrated by examples. A wide range of case studies are tested for demonstrating applicability. Migrating applications into Cloud often requires a major effort in re-designing the users' application source codes. Through using SNC, the users can easily migrate their applications by changing their source codes slightly to delegate the numerical intensive part to the cloud services.

• It is demonstrated that the SNC is suitable for low-end devices, including the embedded devices and mobile devices to efficiently perform numerical intensive computations using the cloud services.

The rest of the paper is organized as follows: related works are presented in Section II. The SNC platform and its key components are presented in Section III. In Section IV, the evaluation method is presented. Experiments are described and results are presented and analyzed in Section V. Discussions are in Section VI and a conclusion is drawn in Section VII.

## II. RELATED WORKS

In this section, we reviewed related works in the fields of the symbolic-numeric computation, JIT compilations and Cloud services.

### A. Symbolic-Numeric Computation

Symbolic-numeric computation is the use of software that combines symbolic and numeric methods to solve problems [23-25]. Symbolic-numeric computation is extensively applied for scientific computation tasks [26-29]. For examples, Amberg el al. [26] generated complete finite element codes in multiple dimensions from a symbolic specification of the mathematical problem in Maple. McPhee et al. [27] combined the symbolic computing methods to dynamic modeling of flexible multibody system. Krowiak [28] employed symbolic computing for determining coefficients in spline-based differential quadrature method. In this, the possibility of defining complicated scientific problems in sufficiently readable syntax allowed researchers to focusing on the scientific problems and trying novel algorithms rapidly, while avoiding program bugs and reducing numerical mistakes. As it has been, symbolic computation is providing an effective program implementation for a wide range of complex scientific computing problems.

Symbolic-numeric computation is widely supported in many popular software systems such as Matlab, Maple, Mathematica, SymPy [30], Theano [31] and SageMath [32]. Matlab, Maple and Mathematica define their own syntax and provide the symbolic toolboxes. SymPy, Theano and SageMath support the symbolic computation in Python-based language.

### B. Just-In-Time Compilation

Usually, work on JIT compilation techniques focuses on the implementation of a specific programming language. In most existing implementations (e.g. Java and C#), JIT is specific to the compilation component in a language. For example, the system is able to collect statistics about how the program is actually running, and it compiles a function which is frequently executed to machine code for direct execution on the hardware. As usual, command line options are provided for a high level control on JIT. However, it is not possible to access the JIT compilation components directly via a programming interface. That is to say, a user cannot directly compile a piece of source code to the machine code at runtime. But, it is likely to generate the intermediate representation (e.g. Java bytecode) and load it into the virtual machine (e.g. JVM) at runtime. Then the virtual machine compiles/translates the bytecode to the native code at certain point. Considering this feature, our SNC platform uses JVM as a backend runtime for symbolic expression evaluation.

As of the year 2000, several projects appear to provide more controls on the JIT compilation through the Application Programming Interfaces (API), including LLVM [28], libJIT and GNU Lightning. These software projects offered a foundation upon which a number of different virtual machines, dynamic scripting languages, or customized rendering routines could be built. In this work, LLVM is used as another backend runtime in which the JIT compiler is used for evaluating the symbolic expressions. Originally, LLVM is developed as a research infrastructure to investigate dynamic compilation techniques for static and dynamic programming languages. Languages including Swift, Rust, Common Lisp, FORTRAN, Haskell, Julia, Objective-C and Lua used LLVM as a backend compiler. LLVM is also an integral part of Apple's latest development tools for Mac OS X and iOS.

The JIT complication [33-36] could significantly increase the flexibility, in comparison with traditional ahead-of-time (AOT) compilation. The common implementation of JIT is to first have AOT compilation of source code to an intermediate representation such as Java bytecode [37] and Microsoft's CIL, and then have JIT compilation to machine code, rather than interpretation of the intermediate representation. In LLVM, the intermediate representation (IR) is designed for using in the three different forms: an in-memory compiler IR; an on-disk bitcode representation which is suitable for fast loading by a JIT compiler; and a human readable assembly language representation. These different forms of LLVM IR are equivalent. The bitcode representation of LLVM IR is used as an intermediate representation for the symbolic expression between the user clients and cloud servers in SNC.

In the existing computer algebra systems (CAS) and general scientific computing languages which support the symbolic manipulations, the result of manipulating a symbolic expression can be numerically evaluated in many different ways. However, none of them uses LLVM to JIT compile the symbol expressions, and all of the ways are inefficient and/or inconvenient in their current implementations. For example, some CAS provided a substitution of symbols to numbers in an expression to yield a numerical value of the expression. This kind of evaluation does not compile the expression when performing evaluation. Some tools provide the so-called JIT compilation for the symbolic expression but they are not real JIT. They use two ways to achieve the numerical evaluation of symbolic expressions. One way is that the symbolic expression is interpreted directly or transformed to an intermediate representation and then an interpreter consumes the intermediate code to perform the evaluation. Examples include the Mathematica's virtual machine, SageMath's interpreter, Matlab's *matlabFunction* function and SymPy's *Lamdify* function. The other way is that the software first generates C/C++/FORTRAN source code, and then it compiles the code by AOT compiler and lastly it links the compiled binary code back to the software environment. Examples include the Mathematica's *compile* function, Theano's *function* function and SymPy's *ufuncify* function. In SNC, we use the new generation of JIT compiler LLVM to achieve the extremely fast in-memory JIT compilation for the symbolic expressions. The highly optimized machine codes run immediately after the compilation. The time of evaluating a symbolic expression is significantly shorter than all of these current implementations.

C. *Cloud-based Services*

Several classical symbolic-numeric tools have been moved to the Cloud environments. Wolfram Cloud [38] is an example of moving Mathematica to the Cloud, and it uses the Wolfram language. SageMath Cloud [39] is the other example of supporting Sage codes and offering cloud service for running SageMath computation online based on Python. Other languages such as Java and C/C++ are not supported directly in the Cloud environment currently.

In this work, we will provide a Cloud platform that supports symbolic manipulation and extremely fast numerical evaluation through the Cloud interfaces in C/C++, Java and Python. Furthermore, a Matlab-syntax like script language - SNC script is designed as an additional way of using SNC. Symbolic manipulations in C++, Java and Python are supported by GiNaC [40], SymJava [41] and SymPy [30, 42] respectively. The fast numerical evaluation is implemented for symbolic expressions by using LLVM compiler infrastructure or Java JVM. In this work, SymJava is a symbolic library developed by us in order to support Java language [41].

III. PLATFORM: CONCEPTS AND IMPLEMENTATIONS

In this section, we present the platform and its component details of SNC, as well as service models. Examples are given to illustrate the workflows of this platform.

A. *Cloud-based Platform*

Generic Cloud platform is illustrated in Fig. 1. As the figure indicated, our platform employs two kinds of Clouds: Compute Cloud that performs the tasks and Data Cloud that manages the data. In Compute Cloud, a PE represents a processing element in which the user task can be executed and a PE can be a virtual machine (VM) in the IaaS solution. In Data Cloud, a DB represents a storage node that stores the user data and results. The DB node can be a VM in the DBaaS solution. In this platform, we separate the task compute and the data storage on different Cloud infrastructures, offering the possibility of choosing a flexible Cloud provider to Cloud users. In Compute Cloud, a task scheduler is built to accept the Internet-based user requests that contain the IR for symbolic expressions; and schedule new tasks to next available PEs. As an assistant to task scheduler, a data scheduler in Data Cloud is built to exchange data with users; store the data; and exchange data with PEs. Fig. 2 outlines the high-level overview of the workflow and functional modules.

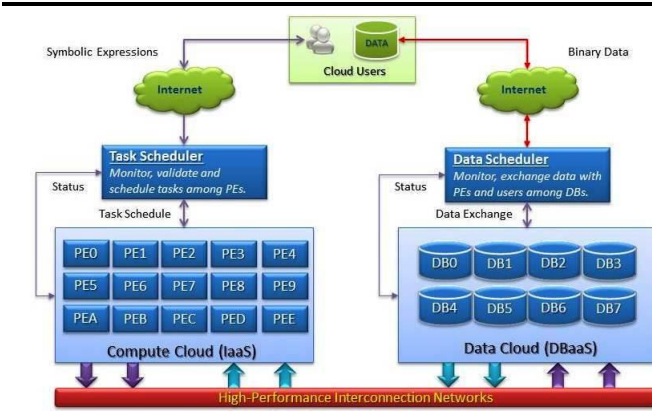

**Fig. 1** Generic Cloud-based platform

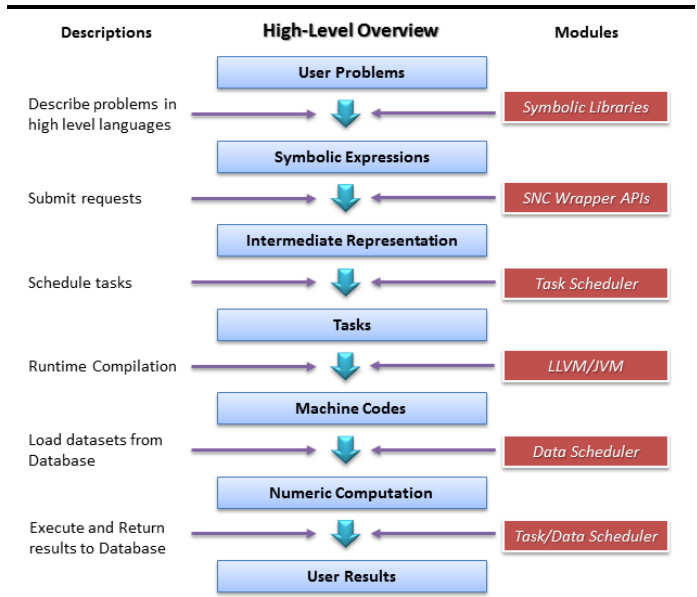

**Fig. 2** High-level overview of workflows and modules

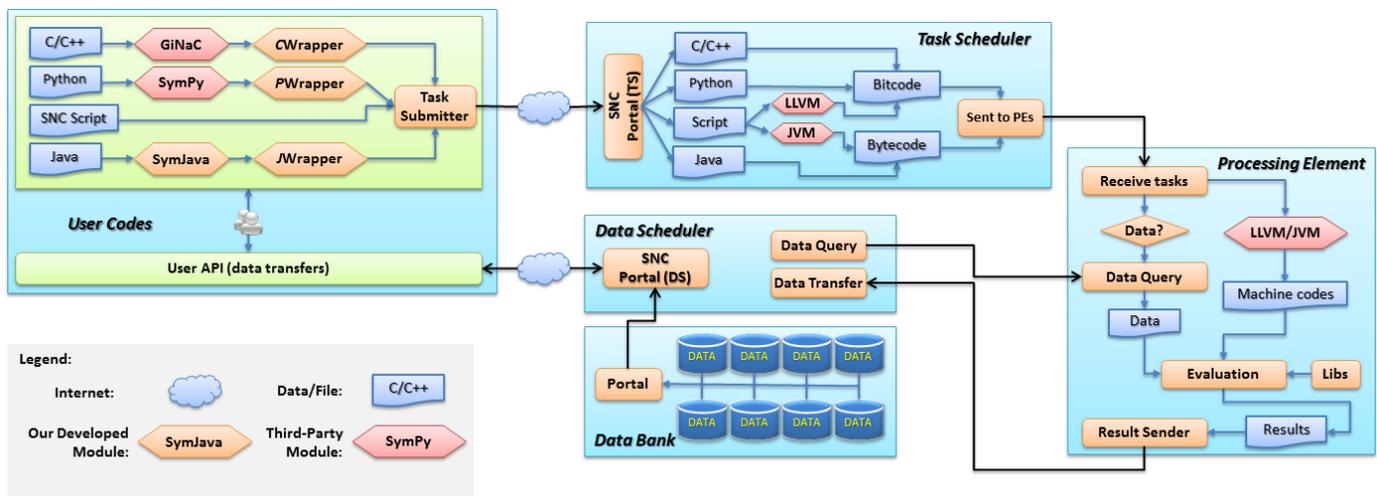

**Fig. 3** SNC components decomposition diagram

### B. Schedulers

Fig. 3 describes the components diagram, at which the colors are used to differentiate whether a module is our developed one or it is adapted from a third party. Five packages are developed: one for user APIs to submit tasks and data, the other two for task and data schedulers respectively, and the last two for offering services on PE and DB nodes respectively.

Task scheduler is a portal. It accepts a user Internet-based request and it parses the symbolic expression that is written in C/C++, Python, Java or SNC script. The processed symbolic expression is combined with appropriate wrapper and complied by LLVM/JVM locally or remotely to yield bitcode/bytecode. Then, it sends the new tasks to the next available PE.

Data scheduler manages the user data. Upon the arrival of a task, PE compiles the bitcode/bytecode to the machine code together with necessary auxiliary libraries. Simultaneously, PE checks whether a user data is needed. If yes, the user data is transferred to the PE with the help of data scheduler. Execution starts as long as the code and data were prepared. The results are stored in DB and meanwhile, an acknowledge message is sent back to the task submitter (the user).

### C. Symbolic Computations

Some dedicated languages such as Maple and Mathematica have their own grammars for symbolic computations. In SNC, the ability to perform symbolic computations in C/C++, Python and Java is supported through third party libraries. Specifically, GiNaC is adapted in SNC for C/C++, SymPy for Python and SymJava for Java. These libraries have similar functions for manipulating symbolic expressions. Unlike Sage or Theano, we do not introduce different classes or functions to define the symbols. The original usage patterns that manipulate symbolic expressions in these libraries are preserved as much as possible. The class and function names are designed as close as possible in all the languages. Therefore, it is easy for users to port their existing codes into SNC in order to take the advantage of the fast numerical evaluation for symbolic expressions. The SNC script supports symbolic manipulations through any of the above-mentioned third party symbolic libraries by using the syntax parsers for the SNC script in a corresponding language.

### D. Just-in-Time Compilation

LLVM compiler infrastructure or Java JVM serves as our symbolic expression compiler in the backend. We demonstrate the compilation process for a symbolic expression here using LLVM. Similar process is for JVM. A symbolic expression is typically stored as an expression tree in GiNac, SymPy and SymJava. An example expression $x*y+z^3$ is shown in Fig 4.

To compile an expression, our current implementation uses a stack to mimic a stack machine to generate the call sequence to the LLVM IRBuilder functions. First, an expression tree is traversed in post-order and a list of atomic expressions can be obtained. For example, the expression x*y+z^3 will result a list [x, y, *, z, 3, ^, +]. Then we feed the list to a pushdown stack to build LLVM IR of the expression. Table 1 lists the pseudo code for this building process.

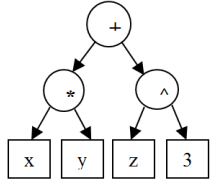

**Fig. 4** The expression tree of x*y+z^3

The generated IR of the example expression x*y+z^3 in LLVM assembly language is shown in Table 2. The optimized machine code of an IR can be just-in-time compiled through using the JIT EnginBuilder in LLVM.

The string representation of an expression in SNC script is parsed by the *parse_expr* function in SymPy and our own developed parsers in C++ and Java based on the Shunting-Yard algorithm. The parsed expression in SNC script is JIT compiled by the same way.

**Table 1:** Pseudo code for generation of IR for an expression with LLVM IRBuilder

```
Input:
    LLVM::IRBuilder irb (reference to a LLVM IRBuilder object)
    LLVM::Function func (reference to a declaration of function in LLVM)
    GiNac::ex expr (reference to a GiNaC object of an expression)
    map<ex, int> argMap (A map from symbols to function arguments of func)
Output:
    LLVM::Value* (The final result of the expression expr)
Algorithm:
Define a local stack: stk
for node in [post-order traverse the expression tree of expr]
{
    case node of
    {
    symbol: stk.push(argMap[node])
    number: stk.push(node)
    summation: //e.g. a+b+c+...
      loop for 'number of add operators' times
        r = s.pop() //right operand
        l = s.pop() //left operand
        stk.push(irb.createFAdd (l, r)) //push float add to stack s
    multiplication: //e.g. a*b*c*...
      loop for 'number of multiply operators' times
        r = stk.pop() //right operand
        l = stk.pop() //left operand
        stk.push(irb.createFMul(l, r) // push float multiply to stack s
    power:
        b = stk.pop() //base
        p = stk.pop() //power
        Function *infun = get declaration of the intrinsic function 'pow'
for b^p (or 'powi' for integer p)
        stk.push(irb.createCall(infun)) //push b^p to stack s
    other GiNac functions: //e.g. sin, cos, ...
        ...
    }
}
return stk.pop() //the final result of expr as the return value
```

**Table 2:** The function of example expression x*y+z^3 in LLVM assembly language

```
define double @myFunc(double, double, double) {
block1:
    %3 = fmul double %0, %1
    %4 = call double @llvm.powi.f64(double %2, i32 3)
    %5 = fadd double %3, %4
    ret double %5
}
```

Compared with ahead-of-time (AOT) compilation, JIT offers better performance since many optimizations are feasible only at runtime. In addition to the optimizations provided by LLVM, more optimizations are conducted in SNC for mathematical formulas. For example, power $b^p$ is optimized to use *llvm.powi* instead of *llvm.pow* when the value of $p$ is an integer despite of the type of $p$.

To provide an easy-to-use interface of JIT compilation for a symbolic expression, wrapper classes are designed in SNC for C/C++, Java and Python, respectively. Four different types of JIT compilation functions are provided to compile the symbolic expression: 1) compile one expression; 2) compile a list of expressions; 3) compile one expression with vectorized arguments; 4) compile a list of expressions with vectorized arguments. Table 3 lists the code segment of the four types of JIT compilation functions in C++.

**Table 3**: Four types of JIT compilation functions in C++

```
typedef double (*JITFunc )(double *args);
typedef int  (*JITBatchFunc)(double *args, double *outAry);
typedef int  (*JITVecFunc )(double **args, double *outAry);
...
class JIT {
public:
// Compile one expression
JITFunc      Compile(vector<string> args, ex &expr);
// Compile a list of expressions
JITBatchFunc BatchCompile(vector<string> args, vector<ex> &exprs);
// Compile one expression with vectorized arguments
JITVecFunc   VecCompile(vector<string> args, size_t nVecLen, ex &expr);
// Compile a list of expressions with vectorized arguments
JITBatchFunc VecBatchCompile(vector<string> args, size_t nVecLen, vector<ex> &exprs);
...
}
```

LLVM C/C++ APIs are used directly in CWrapper for JIT compilation. In Python, llvmpy [43] which is a Python binding for LLVM is used in PWrapper. In Java, two options are offered: (1) an expression is transformed to Java bytecode by using BECL library and then the generated Java bytecode is compiled by Java JIT to machine code [25]; (2) LLVM C API is wrapped through Java JNI interface, then expressions in SymJava can be compiled to bitcode by calling LLVM C API from Java.

E. Example

For the same problem, we present the exemplary codes using different application programming interfaces. The problem is to find out the derivative of the function $R(x, y)$:

$$R(x, y) = 0.127 - \frac{0.194x}{y + 0.194}$$

with respect to *y* and evaluate the derivative at a certain point.

Application programming interfaces for three languages C++, Python and Java are illustrated. Additionally, our SNC Script is used as the fourth way for implementing this example. The advantages of the SNC script will be discussed in the next paragraph. The codes using four different ways are shown in Table 4. In the first three ways, CloudConfig is a class used for choosing a cloud server. The symbols *x* and *y* are the predefined symbol objects. CloudFunc represents a function on the cloud side defined by a given symbolic expression and function arguments. CloudSD (Cloud Shared Data) is a data model for shared data on the cloud. The expression of the function and shared data will be sent to the cloud side through the underlying messaging protocols based on TCP/IP. The evaluation of the function is performed on the cloud. The results are obtained by calling the member function fetchToLocal() of the CloudSD object.

The fourth way, SNC Script is different from the APIs in C++, Python and Java. The script is sent directly to the cloud. The cloud parses the received script and JIT compiles the script to perform computation on the cloud. It should be noted that the installation of the client APIs, the symbolic manipulation libraries and LLVM are not required in this manner.

The ways of using C++, Python or Java APIs provide more advanced symbolic manipulation operations than SNC script in the client side. However, the client side needs to install GiNaC, SymPy or SymJava for support of symbolic manipulation. To do this, the hardware and operating system requirement for the client side is relatively high. The SNC Script way does not require any client side libraries. Thus, for a light-weight device such as an embedded device and low-end mobile device, this way is an option to achieve numerical intensive computations using our SNC Cloud conveniently. This is the advantage of the SNC script. Other advantages, like a webpage-based interface can be easily enabled by using the SNC script. In this manner, the users can access to the SNC cloud easily. Ad-hoc or fast prototype implementation benefit from using a SNC script.

**Table 4:** Example codes for four different ways of using SNC cloud

| SymLLVM: CWrapper | Sympy-llvm: PWrapper | SymJava: JWrapper | SNC Script |
|---|---|---|---|
| CloudConfig.setGlobalTarget("srv1"); <br> ex R=0.127-(x*0.194/(y+0.194)); <br> ex Rdy=R.diff(y); <br> CloudFunc fun=CloudFunc( <br>    lst(x,y), Rdy); <br> CloudSD input=CloudSD("input"); <br> input.init(0.362, 0.556); <br> CloudSD output=CloudSD("output"); <br> fun(output, input); <br> if(output.fetchToLocal()) <br>    cout<<output.getData(0)<<endl; | CloudConfig.setGlobalTarget("srv1") <br> R=0.127-(x*0.194/(y+0.194)) <br> Rdy = R.diff(y) <br> CloudFunc fun = CloudFunc( <br>    [x, y], Rdy) <br> CloudSD input = CloudSD("input").init( <br>    [0.362, 0.556]) <br> CloudSD output=new CloudSD("output") <br> fun.apply(output, input) <br> if(output.fetchToLocal()): <br>    print output.getData(0) | CloudConfig.setGlobalTarget("srv1"); <br> Expr R=0.127-(x*0.194/(y+0.194)); <br> Expr Rdy=R.diff(y); <br> CloudFunc fun=new CloudFunc( <br>    new Expr[]{x,y},Rdy); <br> CloudSD input = new CloudSD().init( <br>    new double[]{0.362, 0.556}); <br> CloudSD output=new CloudSD(); <br> fun.apply(output, input); <br> if(output.fetchToLocal()) <br>   System.out.println(output.getData(0)); | R = 0.127-(x*0.194/(y+0.194)) <br> Rdy=diff(R,y) <br> fun=compile(Rdy) <br> fun(0.362,0.556) |

## IV. EVALUATION METHODOLOGY

We evaluate our approach through implementing the whole SNC design and testing a wide range of symbolic evaluation kernel applications on nowadays popular Cloud providers.

### A. Implementations

User APIs are provided in C/C++, Python and Java, and help the users to submit their tasks and data. Task and data transfer between a client and a server through TCP/IP protocol. For the purpose of efficiency, the APIs for the CWrapper, PWrapper and JWrapper encode the data and IR into different types of messages. The API for the SNC script does not require having any encoded message. The messages between client and server follow a request and response mode. The main types of encoded messages in SNC include: 1) Cloud Shared Data (CloudSD) request and response message; 2) Cloud Function (CloudFunc) request and response message; 3) a general purpose query (CloudQuery) request and response message. The detailed definitions of the main types of messages are listed in Tables 5 and 6. Facility messaging includes the messages for user authentication, node registration, and task scheduling and data management.

On the cloud side, each component runs independently and it communicates with others through TCP/IP protocol. Task and data schedulers run in daemon. In the current implementation, the scheduler follows a straightforward first-in first-service (FIFS) model at scheduling tasks and data storage. PE/DB node registers to its specified task/data scheduler and then demonizes for the next task/data query. For simplicity, PE node takes one single task at a time. DB node indexes its stored files for fast search and it offers the query service. More complex scheduler algorithms and strategies could be investigated in the future but they are not the focus of the present study.

### B. Example

To illustrate the implementation for the process of evaluating a symbolic expression and retrieving the result in SNC, we take the CWrapper code in Table 4 as an example. Fig. 5 shows the client side code, the messaging between client and server, and the corresponding operations on the server side.

First, after performing symbolic manipulation, the derivative *Rdy* of a function *R* is passed to the constructor of class CloudFunc. Symbolic expression *Rdy* is transformed to LLVM IR and packaged into a request message of type CloudFunc. This message is sent to a task scheduler who keeps the message and sends a response to its client.

Second, the input and output data are defined and passed to the function. The input data will be sent to a data scheduler for accessing storage before evaluating the function. An evaluation request CloudQuery (type 1) is then sent to the task scheduler. The task scheduler chooses a free PE and sends the function IR of *Rdy* to this PE. This PE compiles the IR with LLVM JIT compiler and it evaluates the function. Any required data in the function arguments will be fetched from the DB node through the data scheduler.

Last, the returned value of the function will be stored in a DB node and can be fetched to the client side by specifying a name for the returned value which is 'output' in the example.

**Table 5:** Definition of the request messages

| Type | Field | Description |
|---|---|---|
| CloudSD | Magic Flag | Byte='D' |
| | Name Length | Int32 |
| | Data Type | Int32 |
| | Data Length | Int32 (Length in bytes) |
| | Data | Byte[] (CloudSD's name + data) |
| CloudFunc | Magic Flag | Byte='F' |
| | Name Length | Int32 |
| | IR Length | Int32 |
| | Data | Byte[] (Function name + IR) |
| ClouldQuery | Magic Flag | Byte='Q' |
| | Type | Int32=0 (Fetch a CloudSD to local) |
| | | Int32=1 (Evaluate a function) |
| | | Int32=2 (Query machine info) |
| | | Int32=3 (Query installed libraries) |
| | Name Length | Int32 (for the data name or function name) |
| | Returned Name Length | Int32 (for the returned data name) |
| | Data | Byte[] (The function name + the name of returned data) |
| | Argument Numbers | Int32 (for arguments of a function) |
| | Arg1 Name Length | The first argument name of a function |
| | Arg1 Name | |
| | … | … |
| | ArgN Name Length | The last argument name of a function |
| | ArgN Name | |

**Table 6:** Definition of the response messages

| Field | Description |
|---|---|
| Magic Flag | Byte='R' |
| Type | Int32=1 (CloudSDResp) |
| | Int32=2 (CloudFuncResp) |
| | Int32=3 (CloudQueryResp) |
| Status Code | Int32 |
| Name Length | Int32 |
| Message Length | Int32 |
| Data | Byte[] (Name + message) |

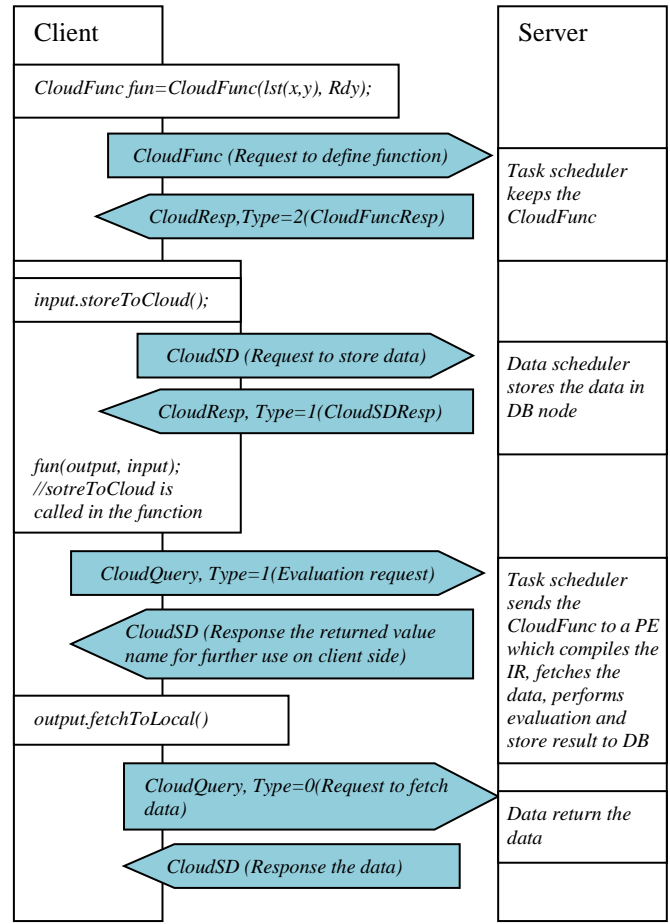

**Fig. 5**: Implementation details of the CWrapper example code in Table 4

### C. Benchmarks

We design two groups of kernel applications to demonstrate the efficacy and advantage of our design and implementation.

First, we show that a JIT-based approach that is used in our implementation is more efficient than other approaches. To this end, we will compare our APIs such as SNC_C++, SNC_Py and SNC_Java with the popular approaches such as Sage 6.5, Theano 0.7, SymPy 0.7.6, Matlab 2015, Mathematica 10 and manually generated C++ code with the O3 option. All the tests are performed on a system with Ubuntu 14.04 LTS installed and hardware configuration is Intel i5-4570 processor at 3.2 GHz with 8GB RAM.

Second, we show that the SNC platform greatly improves the performance for local computing facilities and also supports a wide range of symbolic-numeric computation applications. To this end, we will test four kernel applications, including Monte Carlo (MC) integration, finite element assembly (FEM), nonlinear optimization, and multibody dynamic. These tested applications cover a wide range of today's symbolic-numeric computation.

To perform these tests, we choose three local computing facilities: (1) Raspberry Pi 1 B+; (2) Raspberry Pi 2 B, and (3) Samsung ATIV Book 9 Lite (denoted by 'Samsung Laptop' in Table 5). These local computing facilities are inexpensive and not powerful in computing either. By choosing them, we intend to demonstrate an advantage of cloud computing: providing better performance for demanding applications on lightweight and inexpensive devices. This is important because it shows the performance and capability of our hardware-constrained and software-limited devices can be improved and enhanced using cloud computing. In the capital expenditure, these expensive personal computers would be unnecessary in the Cloud era.

For the cloud, we choose today's most popular Cloud providers, which included (1) Google Compute Engine, (2) Amazon EC2, (3) Microsoft's Windows Azure, (4) Rackspace, (5) HP Helion and (6) VMWare vCloud. Table 5 shows the detail local and cloud hardware configurations.

**Table 5:** System configurations for local computing facilities and the cloud servers

| Environment | Machine Type | Number of CPU Cores and Frequency | Memory (GB) | Geolocation |
|---|---|---|---|---|
| Local | Raspberry Pi 1 B+ | 1 at 700MHz | 0.5 | West US |
| | Raspberry Pi 2 B | 1 at 900MHz | 1 | West US |
| | Samsung Laptop | 4 at 998MHz | 4 | West US |
| Cloud | Google Compute Engine | 1 at 2.50GHz | 3.75 | Central US |
| | AWS EC2 | 1 at 2.50 GHz | 1.0 | West US |
| | Microsoft Azure | 1 at 2.20 GHz | 3.5 | West US |
| | Rackspace | 2 at 2.80 GHz | 3.75 | East US |
| | HP Helion | 2 at 2.40 GHz | 2.0 | West US |
| | VMWare vCloud | 1 at 2.60 GHz | 2.0 | West US |

## V. EXPERIMENTS

### A. Efficiency of JIT-based Implementations

We select two typical examples for numerical evaluation of symbolic expressions: (1) Taylor expansion and (2) polynomial with fractional powers. The configurations are:

1) Evaluation of Taylor expansion for $e^x$ at $x = 0$,
$$e^x \approx \sum_{n=0}^{N} \frac{x^n}{n!} \text{ where } N = 0, \ldots, 9.$$
2) Polynomial with fractional powers,
$$f(x) = \sum_{n=1}^{N} \sqrt[n]{x} \text{ where } N = 1, \ldots, 9.$$

Figs. 6 and 7 show the evaluation time in seconds vs. the problem complexity N. CPU time in seconds is used for timing.

The evaluation time is presented in a log scale with base 10 for purpose of clarity. Clearly, the results show that our JIT based implementations (SNC_C++, SNC_Py and SNC_Java) have greatly outperformed the popular approaches that are still based on the interpretation and/or source code generation techniques. Thus, the JIT-based approach is now the most efficient way for fast numerical evaluation of symbolic expressions. It should be noted that SNC_C++ is faster than C++_O3 in Fig. 6. The reason is that we optimized the computation $x^n$ by choosing an efficient algorithm for integer value of $n$ at runtime ($n$ could be declared as double) while C++_O3 doesn't provide the optimization on such level since C++ compiler merely has the static information of the variables.

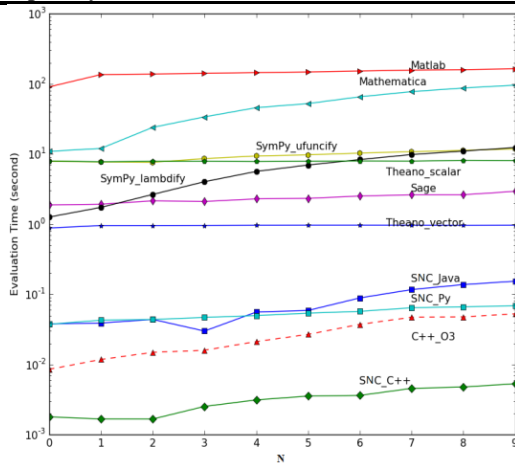

**Fig. 6** Evaluation time (in seconds) of Taylor expansions

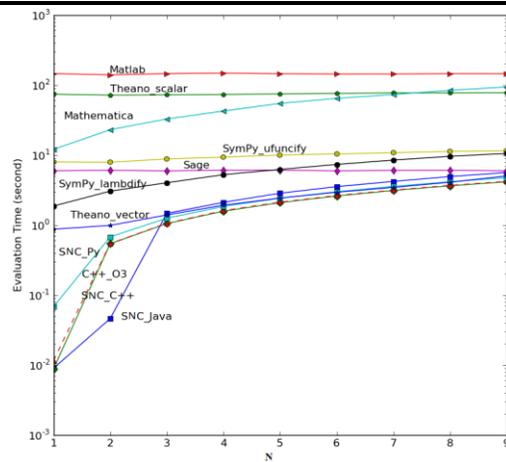

**Fig. 7** Evaluation time (in seconds) of polynomial with fractional powers

### B. Performing Computation on the Cloud

Four symbolic-numeric computation applications are tested: (1) Monte Carlo numerical integration, (2) finite element assembly, (3) nonlinear minimization solver and (4) multibody dynamics for a double pendulum. We perform these tests on both local computing facilities (if local hardware permits) and the cloud servers (as in Table 5). We use the wall clock time in seconds to measure the performance. For the cloud-based tests, the results included both the symbolic-numeric computation and the client-server communication time.

*Configurations*

The configurations are as follows.

*1) Monte Carlo (MC) numerical integration*

$$I = \int_\Omega \sin\left(\sqrt{\log(x+y+1)}\right) d\Omega$$

where $\Omega = \left\{(x,y) \mid \left(\frac{1}{8}\right)^2 \leq \left(x-\frac{1}{2}\right)^2 + \left(y-\frac{1}{2}\right)^2 \leq \left(\frac{1}{4}\right)^2\right\} \cup \left\{(x,y) \mid \left(\frac{3}{8}\right)^2 \leq \left(x-\frac{1}{2}\right)^2 + \left(y-\frac{1}{2}\right)^2 \leq \left(\frac{1}{2}\right)^2\right\}$

Fig. 8 shows the integration domain $\Omega$ where the red dots represent the points in the domain $\Omega$. In the test, the cloud client transforms the expression of the integration to an IR. The IR and the required number of random points are packaged and sent to a cloud scheduler for computing the integration. The scheduler forwards the request package to an idle PE. The PE further compiles the IR to machine. By running the machine code the PE generates random points in the square [0,1]×[0,1] and performs Monte Carlo numerical integration.

*2) Finite Element (FE) assembly*

This example is for solving the Poisson's equation with zero boundary values and a nonzero right hand side with finite element method (FEM):

$$\Delta u = -4 \times (x^4 + y^4) \text{ in } \Omega,$$
$$u(x,y) = 0 \text{ on } \partial\Omega.$$

where $\Omega = \{(x,y) \mid |x| \leq 3, |y| \leq 3\}$, $\partial\Omega$ is the boundary of $\Omega$.

Fig. 9 shows a random mesh of the domain $\Omega$. In the test, the cloud client transforms several symbolic expressions in the weak form of the Poisson's equation to an IR of the expressions. The IR together with an array of random numbers are packaged and sent to the cloud scheduler. The scheduler finds an idle PE to run the assembly task which includes the evaluation of several expressions. The PE generates a random mesh on $\Omega$ based on the given random numbers and performs finite element assembly on the random mesh. The Java API is used in tests (1) and (2).

*3) Nonlinear minimization problem*

This example considers finding the global minimum of Griewank function (Fig. 10)

$$f(\vec{x}) = \sum_{i=1}^{d} \frac{x_i^2}{4000} - \prod_{i=1}^{d} \cos\left(\frac{x_i}{\sqrt{i}}\right) + 1.$$

We symbolically define the object function using the C++ API. Then the cloud client sends the symbolic expression of Griewank function to the cloud scheduler. After computing derivatives and compiling the expressions of the derivatives, the scheduler finds a PE to run the task. Nlopt library is used on PE to find the global minimum of Griewank function. The dimension $d$ of the function in the experiment ranges from 10 to 150.

*4) Multibody dynamics for a double pendulum*

This example is chosen from PyDy project [44]. A double pendulum is a pendulum with another pendulum attached to its end. The motion of a double pendulum is governed by four coupled first order ordinary differential equations (ODE). For certain energies its motion is chaotic. In PyDy, the ODE is obtained through symbolic computation by SymPy mechanics package. PyDy provides several ways to solve the resulting system of ODEs. The C++, Python or Matlab source codes for solving the system of ODEs can be generated by PyDy. Instead of using the way of source code generation, we use SNC python API. The resulting system of ODEs is sent to the cloud and it is solved efficiently by using JIT compilation on the cloud side. Fig. 11 shows the solution of the system of ODEs.

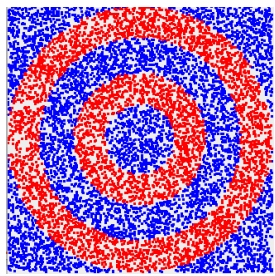 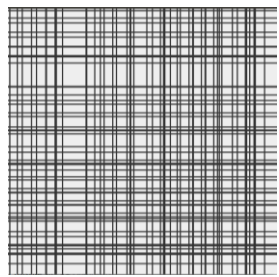 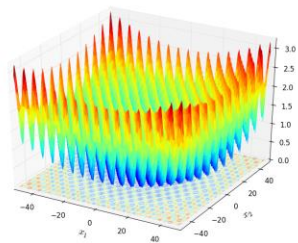 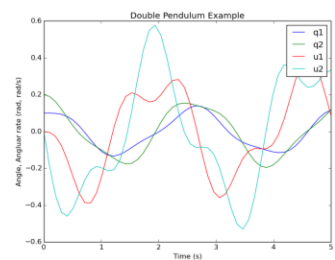

**Fig. 8** MC integration domain      **Fig. 9** FEM random mesh      **Fig. 10** Griewank function      **Fig. 11** Solution of a double pendulum

*Performance Results*

The evaluation time is presented in a log scale with base 2 for purpose of clarity. The Monte Carlo results are shown in Figs. 12 to 14. The results show that for a small number of random points (for example, less than 1.0E+5) the local evaluation is faster than the cloud side evaluation. The computational complexity for the Monte Carlo example is proportional to the amount of data transferred (the number of random points). By inspecting the trend of time used, we see that with increasing of the number of random points, the cloud side evaluation is dominantly faster than the local evaluation. For example, it is approximately 16 times faster than the local evaluation (Raspberry Pi Model 1 and 2) in the case of 1.0E+8 points on the cloud side (excluding HP Helion and will be explained later).

The finite element results are shown in Figs. 15 to 17. The results show that even for a small sized problem the cloud side evaluation is much faster than the local evaluation. The reason is that the task is computing intensive and its computational complexity is quadratic with respect to the number of data transferred. In addition, the hardware on the cloud is much superior to the local computing facilities.

It is clear that the performance of the cloud side evaluation depends on the configuration of the provision servers. In this finite element assembly test, the HP Helion instance is worse than the local evaluation on the laptop due to low effective CPU frequency of the virtual machine.

The nonlinear minimization results are shown in Figs. 18 to 20. The multibody dynamics results are shown in Figs. 21 to 23. Raspberry Pi systems failed due to its limited memory capacity and CPU capability. However, through using the SNC, even the very low-end computing facilities "performed" these time- and memory-consuming tasks successfully - this has demonstrated the significance of the cloud services the SNC provided. Again, the SNC greatly outperformed the standalone Samsung laptop system. In these tests, the users only need to simply describe the symbolic expressions and submit them to the SNC. The SNC automated a complicated symbolic computation and numerical evaluation. This fully implemented the concept of providing the efficient symbolic-numeric computation services on the Cloud.

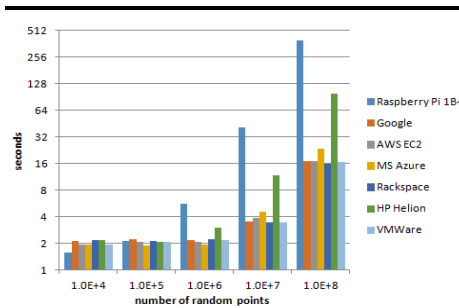
**Fig. 12** Monte Carlo Integration on Raspberry Pi Model 1 B+

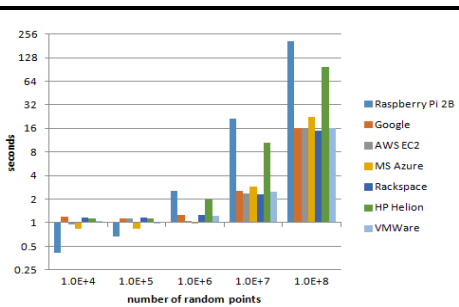
**Fig. 13** Monte Carlo Integration on Raspberry Pi Model 2 B

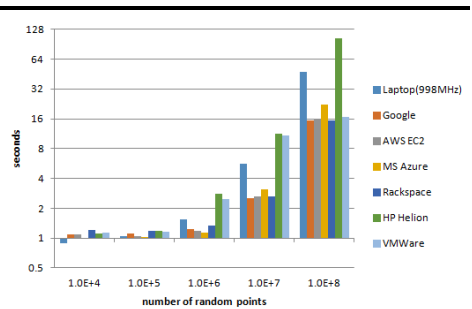
**Fig. 14** Monte Carlo Integration on Samsung Laptop

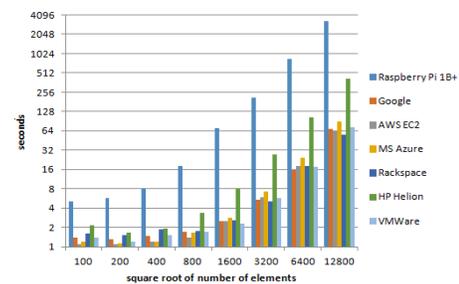
**Fig. 15** Finite Element Assembly on Raspberry Pi Model 1 B+

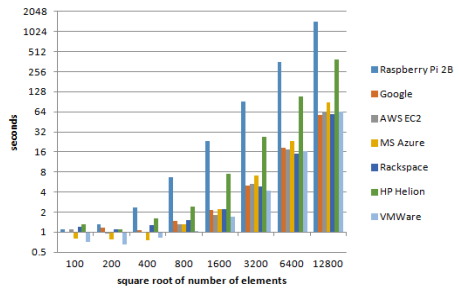
**Fig. 16** Finite Element Assembly on Raspberry Pi Model 2 B

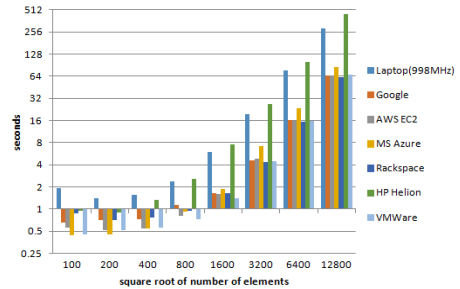
**Fig. 17** Finite Element Assembly on Samsung Laptop

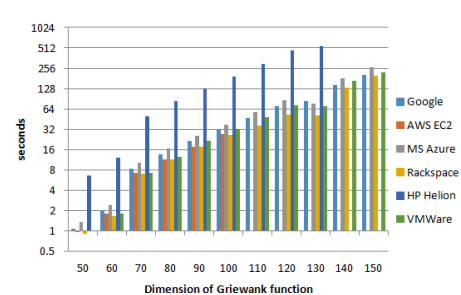
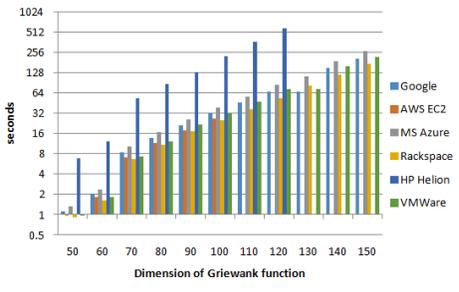
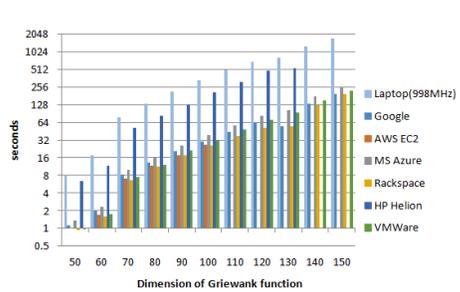

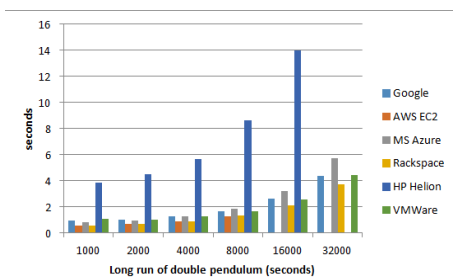

**Fig. 18** Nonlinear minimization on Raspberry Pi Model 1 B+

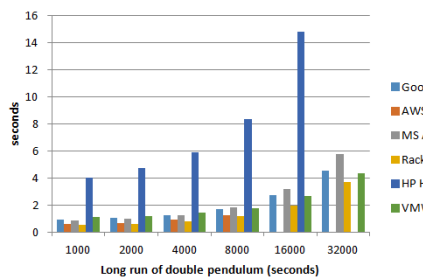

**Fig. 19** Nonlinear minimization on Raspberry Pi Model 2 B

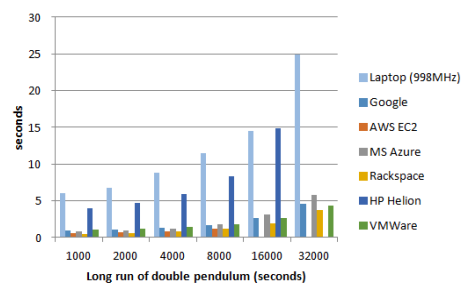

**Fig. 20** Nonlinear minimization on Samsung Laptop

**Fig. 21** Multibody dynamics on Raspberry Pi Model 1 B+

**Fig. 22** Multibody dynamics on Raspberry Pi Model 2 B

**Fig. 23** Multibody dynamics on Samsung Laptop

## VI. DISCUSSIONS AND FUTURE WORKS

The multi-language support and cross-platform capabilities are demonstrated by the real-world applications. Through using the SNC, the users easily delegate the numerical intensive part to the cloud services. The performance is greatly improved.

However, in the case of certain applications with small sized computation tasks, the local computation may be faster than the client-server communication in SNC. Under this circumstance, performing local computation is a straightforward choice. To resolve this issue, we designed a threshold option at the client side. This option allows a user to decide the threshold value based on a problem size. Only computation tasks that exceeded the threshold can be submitted to the SNC cloud; otherwise, the local computation will be conducted.

The portability of SNC is considered and enhanced in two aspects: (1) the user client-cloud server separation and (2) the task scheduler-processing element separation. The client-server communication follows a stipulated transmission protocol. In this, we implemented this protocol in Section IV. An alternative choice for this implementation could be based on some open source projects like Apache Thrift or Google Protocol Buffer that support multi-language and multi-platform communication in a client-server mode. The task scheduler-process element communication follows the intermediate representation (IR) such as LLVM IR and Java bytecode. Two backend systems are provided: LLVM and JVM. For the LLVM system, the task scheduler and process element are separated by LLVM IR. The process elements are the main computation unit of the SNC. By using the LLVM target-independent code generator, the LLVM IR is translated to the machine code for a specified target (e.g. CPU or GPU) in a binary machine code. For the JVM system, JVM is well known as 'write-once run-anywhere'. Together, the portability of the SNC architecture on heterogeneous cloud is enhanced by the portability of the backend systems and the client-server design.

In the future, the SNC platform should be further extended by adding a cloud security solution, being adapted for certain domain-specific application frameworks and supporting the large scale matrix-based operations. These works could extend the usability of the SNC cloud service platform in the real word.

## VII. CONCLUSIONS

In this paper, we presented a cloud service platform for symbolic-numeric computation, the SNC platform. The SNC supported popular computer languages, such as C/C++, Python, Java, for the symbolic-numeric computation with Just-In-Time (JIT) compilation. The platform also integrated cloud services through the designed modules and communication protocols. Thus, the SNC supports a wide range of languages and has the cross-platform interoperability so it is superior to other existing platforms such as Wolfram Cloud and SageMath.

By experimenting a number of user applications on popular cloud providers (such as Google Compute Engine, Amazon EC2, Microsoft Azure, Rackspace, HP Helion and VMWare vCloud), we demonstrated that (a) the SNC platform speeds up the symbolic-numeric computation significantly (2~16 times faster compared with these local computations); and (b) this cloud-based service platform enables lightweight devices or mobile devices to perform numerical intensive computations.

In the implementation, the SNC provided the intuitive user interfaces and syntax, and supported the symbolic expression in a human-readable format. This minimized the syntax difference among different computer program languages. This effort helps to lower the difficulty of learning the platform to program.

As Cloud computing is raising increasing attention, the SNC platform, as the first enabler, will help the research community to smoothly adopt the cloud computing technique. This offered a way to stimulate the need for using the cloud computing for the symbolic-numeric computation in the field of scientific research.

### REFERENCES

[1] L. M. Vaquero, L. Rodero-Merino, J. Caceres, and M. Lindner, "A break in the clouds: towards a cloud definition," *SIGCOMM Comput. Commun. Rev.,* vol. 39, pp. 50-55, 2008.

[2] M. Armbrust, A. Fox, R. Griffith, A. D. Joseph, R. Katz, A. Konwinski, *et al.*, "A View of Cloud Computing," *Communications of the ACM,* vol. 53, pp. 50-58, Apr 2010.

[3] P. Mell and T. Grance, "The NIST Definition of Cloud Computing," *Communications of the ACM,* vol. 53, pp. 50-50, Jun 2010.
[4] S. Hazelhurst, "Scientific computing using virtual high-performance computing: a case study using the Amazon elastic computing cloud," in *Proceedings of the 2008 annual research conference of the South African Institute of Computer Scientists and Information Technologists on IT research in developing countries: riding the wave of technology*, 2008, pp. 94-103.
[5] L. Z. Wang, J. Tao, M. Kunze, A. C. Castellanos, D. Kramer, and W. Karl, "Scientific Cloud Computing: Early Definition and Experience," *Hpcc 2008: 10th Ieee International Conference on High Performance Computing and Communications, Proceedings,* pp. 825-830, 2008.
[6] C. Evangelinos and C. Hill, "Cloud computing for parallel scientific HPC applications: Feasibility of running coupled atmosphere-ocean climate models on Amazon's EC2," *ratio,* vol. 2, pp. 2-34, 2008.
[7] C. Vecchiola, S. Pandey, and R. Buyya, "High-Performance Cloud Computing: A View of Scientific Applications," *2009 10th International Symposium on Pervasive Systems, Algorithms, and Networks (Ispan 2009),* pp. 4-16, 2009.
[8] R. Buyya, C. S. Yeo, S. Venugopal, J. Broberg, and I. Brandic, "Cloud computing and emerging IT platforms: Vision, hype, and reality for delivering computing as the 5th utility," *Future Generation Computer Systems,* vol. 25, pp. 599-616, Jun 2009.
[9] S. Ostermann, A. Iosup, N. Yigitbasi, R. Prodan, T. Fahringer, and D. Epema, "A Performance Analysis of EC2 Cloud Computing Services for Scientific Computing," *Cloud Computing,* vol. 34, pp. 115-131, 2010.
[10] J. J. Rehr, F. D. Vila, J. P. Gardner, L. Svec, and M. Prange, "Scientific computing in the cloud," *Computing in Science & Engineering,* vol. 12, pp. 34-43, 2010.
[11] S. Srirama, O. Batrashev, and E. Vainikko, "SciCloud: scientific computing on the cloud," in *Proceedings of the 2010 10th IEEE/ACM International Conference on Cluster, Cloud and Grid Computing*, 2010, pp. 579-580.
[12] A. Iosup, S. Ostermann, M. N. Yigitbasi, R. Prodan, T. Fahringer, and D. H. Epema, "Performance analysis of cloud computing services for many-tasks scientific computing," *Parallel and Distributed Systems, IEEE Transactions on,* vol. 22, pp. 931-945, 2011.
[13] Y. Deng, P. Zhang, C. Marques, R. Powell, and L. Zhang, "Analysis of Linpack and power efficiencies of the world's TOP500 supercomputers," *Parallel Computing,* vol. 39, pp. 271-279, 2013.
[14] C. Moler and P. J. Costa, "MATLAB Symbolic Math Toolbox," *User's Guide, Version,* vol. 2, pp. 01760-1500, 1997.
[15] K. O. Geddes and G. J. Fee, "Hybrid symbolic-numeric integration in MAPLE," in *Papers from the international symposium on Symbolic and algebraic computation*, 1992, pp. 36-41.
[16] S. Wolfram, "The Mathematica Book," *Cambridge University Press and Wolfram Research, Inc., New York, NY, USA and,* vol. 100, pp. 61820-7237, 2000.
[17] M. K. J Su, Y Liu, Z Lin, N Pantong, H Liu, "Optical imaging of phantoms from real data by an approximately globally convergent inverse algorithm," *Inverse problems in science and engineering,* vol. 21, p. 26, 2013.
[18] N. P. A Rhoden, Y Liu, J Su, H Liu, "A globally convergent numerical method for coefficient inverse problems with time-dependent data," *Applied Inverse Problems,* p. 24, 2013.
[19] J. S. Y Liu, ZJ Lin, S Teng, A Rhoden, N Pantong, H Liu, "Reconstructions for continuous-wave diffuse optical tomography by a globally convergent method," *Journal of Applied Mathematics and Physics,* vol. 2, p. 10, 2014.
[20] D. A. H Patrick E. Farrell, Simon W. Funke and Marie E. Rognes, "Automated derivation of the adjoint of high-level transient finite element programs," *SIAM Journal on Scientific Computing,* vol. 35, p. 25, 2013.
[21] J. McCarthy, "Recursive functions of symbolic expressions and their computation by machine, Part I," *Communications of the ACM,* vol. 3, pp. 184-195, 1960.
[22] Y. Liu, P. Zhang, and M. Qiu, "Fast Numerical Evaluation for Symbolic Expressions in Java," in *2015 IEEE 17th International Conference on High Performance Computing and Communications, 2015 IEEE 7th International Symposium on Cyberspace Safety and Security, and 2015 IEEE 12th International Conference on Embedded Software and Systems*, 2015, pp. 599-604.
[23] J. Grabmeier, E. Kaltofen, and V. Weispfenning, *Computer Algebra Handbook: Foundations, Applications, Systems* vol. 1: Springer Science & Business Media, 2003.
[24] D. Wang and L.-H. Zhi, *Symbolic-Numeric Computation*: Springer Science & Business Media, 2007.
[25] U. Langer and P. Paule, *Numerical and Symbolic Scientific Computing: Progress and Prospects*: Springer Science & Business Media, 2011.
[26] G. Amberg, R. Tonhardt, and C. Winkler, "Finite element simulations using symbolic computing," *Mathematics and Computers in Simulation,* vol. 49, pp. 257-274, Sep 1999.
[27] J. McPhee, C. Schmitke, and S. Redmond, "Dynamic modelling of mechatronic multibody systems with symbolic computing and linear graph theory," *Mathematical and Computer Modelling of Dynamical Systems,* vol. 10, pp. 1-23, Mar 2004.
[28] A. Krowiak, "Symbolic computing in spline‐based differential quadrature method," *Communications in Numerical Methods in Engineering,* vol. 22, pp. 1097-1107, 2006.
[29] B. Harvey, *Computer science logo style: Symbolic computing* vol. 1: MIT press, 1997.
[30] D. Joyner, O. Certik, A. Meurer, and B. E. Granger, "Open source computer algebra systems: SymPy," *ACM Commun. Comput. Algebra,* vol. 45, pp. 225-234, 2012.
[31] J. Bergstra, O. Breuleux, F. Bastien, P. Lamblin, R. Pascanu, G. Desjardins*, et al.*, "Theano: a CPU and GPU math expression compiler," in *Proceedings of the Python for scientific computing conference (SciPy)*, 2010, p. 3.
[32] W. Stein, "Sage: Open Source Mathematical Software," ed: The Sage Group, 2008.
[33] K. Ishizaki, M. Kawahito, T. Yasue, M. Takeuchi, T. Ogasawara, T. Suganuma*, et al.*, "Design, implementation, and evaluation of optimizations in a Java (TM) Just-in-Time compiler," *Concurrency-Practice and Experience,* vol. 12, pp. 457-475, May 2000.
[34] C. Lattner and V. Adve, "LLVM: A compilation framework for lifelong program analysis & transformation," in *Code Generation and Optimization, 2004. CGO 2004. International Symposium on*, 2004, pp. 75-86.
[35] P. Bonzini and L. Michel, "GNU Lightning library," ed, 2004.
[36] G. Barany, "pylibjit: A JIT Compiler Library for Python," in *Software Engineering (Workshops)*, 2014, pp. 213-224.
[37] J. Gosling, "Java Intermediate Bytecodes - Acm Sigplan Workshop on Intermediate Representations (Ir 95)," *Sigplan Notices,* vol. 30, pp. 111-118, Mar 1995.
[38] *Wolfram Cloud*. Available: https://www.wolframcloud.com/
[39] *SageMath Cloud*. Available: https://cloud.sagemath.com/
[40] C. Bauer, A. Frink, and R. Kreckel, "Introduction to the GiNaC framework for symbolic computation within the C++ programming language," *Journal of Symbolic Computation,* vol. 33, pp. 1-12, Jan 2002.
[41] Y. Liu, P. Zhang, and M. Qiu, "Fast Numerical Evaluation for Symbolic Expressions in Java," presented at the 17th IEEE International Conference on High Performance and Communications (HPCC 2015), 2015.
[42] O. Certik, "SymPy Python library for symbolic mathematics," Technical report (since 2006), http://code.google.com/p/sympy/ (accessed November 2009).
[43] *llvmpy*. Available: https://github.com/llvmpy
[44] G. Gede, D. L. Peterson, A. S. Nanjangud, J. K. Moore, and M. Hubbard, "Constrained multibody dynamics with Python: From symbolic equation generation to publication," in *ASME 2013 International Design Engineering Technical Conferences and Computers and Information in Engineering Conference*, 2013, pp. V07BT10A051-V07BT10A051.

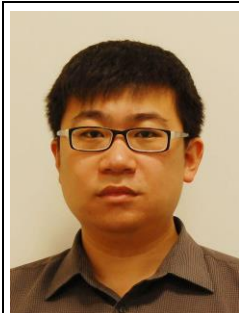

**Peng Zhang** received the BS degree in mathematics from Nankai University in 2003 and the MS degree in parallel computing from Nankai Institute of Scientific Computing in 2006, and the PhD degree in applied mathematics from Stony Brook University, New York, USA in 2012. Currently, he is a Senior Research Associate at Stony Brook University, NY, USA. Dr. Zhang's research interests include development and enhancement of models, algorithms, software and problem-solving environments for domain-specific applications that reply on the high-performance computing (HPC) technologies. His work has appeared in over 50 publications and presentations. He is the member of IEEE.

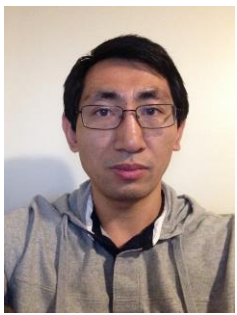

**Yueming Liu** received his B.S. degree in information and computing science and Ph.D degree in computational mathematics from Nankai University. He was a senior software engineer at Qunar.com. He worked as a faculty research associate at the university of Texas at Arlington. He worked at Spokeo.com as a software engineer and now is a senior software engineer in OpenX technologies, Inc. His main academic and industry interests are in Finite Element methods, PDE inverse problems, high performance computing and bytecode/bitcode oriented software development paradigm with Just-in-Time compilation.

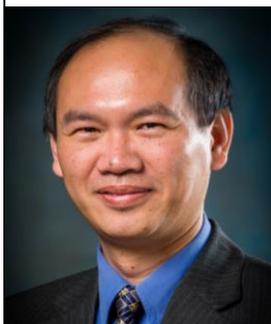

**Meikang Qiu** received the BE and ME degrees from Shanghai Jiao Tong University and received Ph.D. degree of Computer Science from University of Texas at Dallas. Currently, he is an Adjunct Professor at Columbia University and Associate Professor of Computer Science at Pace University. He is an IEEE Senior member and ACM Senior member. He is the Chair of IEEE Smart Computing Technical Committee. His research interests include cyber security, cloud computing, big data storage, hybrid memory, heterogeneous systems, embedded systems, operating systems, optimization, intelligent systems, sensor networks, etc.